\documentclass[preprintnumbers,amsmath,amssymb,showpacs]{revtex4}

\usepackage{bm}
\makeatletter
\newcommand{\rmnum}[1]{\romannumeral #1}
\newcommand{\Rmnum}[1]{\expandafter\@slowromancap\romannumeral #1@}
\makeatother

\begin{document}

\title{Monogamy inequality in terms of entanglement measures based on distance for pure multiqubit states}

\author{Limin Gao}
\author{Fengli Yan}
\email{flyan@hebtu.edu.cn}
\affiliation {College of Physics, Hebei
Normal University, Shijiazhuang 050024, China}
\author{Ting Gao}
\email{gaoting@hebtu.edu.cn}
\affiliation {School of Mathematical Science, Hebei
Normal University, Shijiazhuang 050024, China}

\date{\today}

\begin{abstract}
Using very general arguments, we  prove that any entanglement measures based on distance must be maximal on pure states. Furthermore, we show that Bures measure of entanglement and geometric measure of entanglement satisfy the monogamy inequality on all pure multiqubit states. Finally, using the power of Bures measure of entanglement and geometric measure of entanglement, we  present a class of tight monogamy relations for pure states of multiqubit systems.
\end{abstract}

\pacs{03.67.Mn, 03.65.Ud, 03.67.-a}

\maketitle

Entanglement is one of the most important features present in quantum theory. An important property distinguishing   entanglement  from classical correlations  is the monogamy of entanglement (MOE) [1, 2], which means that a quantum subsystem in a multipartite quantum system entangled with another subsystem limits its entanglement with the remaining ones.  It means that entanglement cannot be freely shared unconditionally among the multipartite quantum systems. For instance, for a three partite quantum system \emph{A}, \emph{B} and \emph{C}, if \emph{A} and \emph{B} share  maximal entanglement, then they share no entanglement with  \emph{C}.  MOE indicates that there is a trade-off on the amount of entanglement between the pairs \emph{AB} and \emph{AC}.

MOE is very  important in the context of quantum cryptography because  it restricts on the amount of information that an eavesdropper could potentially obtain about the secret key extraction. As a matter of fact, many information-theoretic protocols [3-5] can be guaranteed secure by the constraints on the sharing of entanglement.

In 2000 Coffman, Kundu, and Wootters proved the first mathematical characterization of MOE for three-qubit state in terms of squared concurrence, known as CKW-inequality [1]. Osborne and Verstraete generalized  this inequality to arbitrary multiqubit systems [6]. Later, it was proved that the same monogamy inequalities hold for other entanglement measures [7-13]. These monogamy relations play a very  important role in quantum information theory [14], condensed-matter physics [15] and even black-hole physics [16].

One  class of entanglement measures based on distance was  proposed in [17, 18]. Those measures quantify the minimum distance between a given state and the separable states. Examples of such measures are the Bures measure of entanglement [18] and geometric measure of entanglement [19], which are the widely used entanglement measures in the multiparticle system.

In this paper, we focus on the entanglement measures based on distance. First, using very general arguments we show that any entanglement measures based on distance must be maximal on pure states. We then prove that the Bures measure of entanglement and geometric measure of entanglement, as a special case of entanglement measures based on distance, satisfy the monogamy inequality for multiqubit pure states. It is well known that tightening the monogamy relations can provide a precise characterization of the entanglement in multipartite systems. So we also establish a class of tight monogamy relations for pure states of multiqubit systems by using the power of Bures measure of entanglement and geometric measure of entanglement.

We first present some notations and definitions. Consider a quantum system $AB$ consisted of subsystems $A$ and $B$. Let $\mathcal{H}_{A}$ and $\mathcal{H}_{B}$ be two finite dimension Hilbert spaces corresponding to the subsystems $A$ and $B$ respectively. A pure state $|\psi_{AB}\rangle$ of the quantum system $AB$ in the tensor product space $\mathcal{H}_{A}\otimes \mathcal{H}_{B}$ is said to be separable, if it can be written in the form $|\psi_{AB}\rangle=|\varphi_{A}\rangle\otimes |\varphi_{B}\rangle$, where $|\varphi_{A}\rangle\in \mathcal{H}_{A}$ and $|\varphi_{B}\rangle\in \mathcal{H}_{B}$.  A mixed state $\sigma_{AB}$ of the quantum system $A,B$  is called separable, if there is a probability distribution $\{p_{i}\}$ such that
\begin{equation}\label{1}
\begin{aligned}
 \sigma_{AB}= \sum_{i}p_{i} |\varphi^{i}_{A}\rangle\langle\varphi^{i}_{A}|\otimes|\varphi^{i}_{B}\rangle\langle\varphi^{i}_{B}|
\end{aligned}
\end{equation}
holds. We use $S$ to denote the set of separable states, which is a convex set and its extreme points are  pure states.

The entanglement measure based on distance for a state $\rho_{AB}$ of the quantum system $A,B$ is  defined as [17, 18]
\begin{equation}\label{2}
\begin{aligned}
 E_{A|B}(\rho_{AB})= \min_{\sigma\in S} D(\rho_{AB},\sigma_{AB}),
\end{aligned}
\end{equation}
where the minimum is taken over all separable states, and $D$ is any measure of distance between the two density matrices $\rho_{AB}$ and $\sigma_{AB}$ in $S$ such that $E_{A|B}(\rho_{AB})$ satisfies the following general properties [17, 18].

(\rmnum{1}) $E_{A|B}(\rho_{AB})\geq 0$. $E_{A|B}(\rho_{AB})=0$ iff $\rho_{AB}$ is separable.

(\rmnum{2}) $E_{A|B}(\rho_{AB})$ is invariant under local unitary transformation, i.e.,
\begin{equation}\label{3}
E_{A|B}(\rho_{AB})=E_{A|B}(U_{A}\otimes U_{B}\rho_{AB}U_{A}^\dagger\otimes U_{B}^\dagger).
\end{equation}
Here $U_{A}$ and $U_{B}$ are unitary operations acting on subsystems $A$ and $B$ respectively.

(\rmnum{3}) $E_{A|B}(\rho_{AB})$ is non-increasing on average under local operations and classical communication (LOCC). That is, if a LOCC protocol applied to state $\rho_{AB}$, the state $\varrho^{i}_{AB}$ with label $i$ is obtained with probability $p_{i}$, then
\begin{equation}\label{4}
E_{A|B}(\rho_{AB})\geq \sum_{i}p_{i}E_{A|B}(\varrho^{i}_{AB}).
\end{equation}

(\rmnum{4}) $E_{A|B}(\rho_{AB})$ is invariant under attaching a local ancilla, i.e.,
\begin{equation}\label{5}
E_{A|B}(\rho_{AB})=E_{A|BC}(\rho_{AB}\otimes |i\rangle\langle i|_{C}),
\end{equation}
where \{$|i\rangle$\} is  a quantum  state of the ancilla $C$.

Let $\rho_{ABC}$ be a tripartite state in a finite dimensional Hilbert space $\mathcal H_{A}\otimes \mathcal H_{B}\otimes \mathcal H_{C}$, and $\rho_{A|BC}$ denote the state $\rho_{ABC}$ viewed as a bipartite state with partitions $A$ and $BC$. $E_{A|BC}(\rho_{A|BC})$ cannot increase upon tracing out subsystems, i.e.
\begin{equation}\label{6}
E_{A|BC}(\rho_{A|BC})\geq E_{A|B}(\rho_{AB}),
\end{equation}
where $\rho_{AB}=\text{tr}_{C}(\rho_{ABC})$. Indeed, Eq. (6) is a special case of Eq. (4) since the partial trace is a special LOCC.

Now we can prove the following theorem.

\textbf{Theorem 1.}  The entanglement measure  based on distance, $E(\rho_{AB})$, which satisfies Eqs. (3), (5) and (6), is non-increasing under operations on one side, that is $E_{A|B}[\Lambda_B(\rho_{AB})]\leq E_{A|B}(\rho_{AB})$, where $\Lambda_B(\rho_{AB})=\text{tr}_{C}(U_{BC}\rho_{AB}\otimes|i\rangle\langle i|_{C}U_{BC}^\dag).$

\emph{Proof}. Since $E$ satisfy Eqs. (3), (5) and (6), one finds
\begin{equation}\label{7}
\begin{aligned}
E_{A|B}(\rho_{AB})= & \ E_{A|BC}(\rho_{AB}\otimes|i\rangle\langle i|_{C})\\
= &\ E_{A|BC}(U_{BC}\rho_{AB}\otimes|i\rangle\langle i|_{C}U_{BC}^\dag)\\
\geq &\ E_{A|B}[\text{tr}_{C}(U_{BC}\rho_{AB}\otimes|i\rangle\langle i|_{C}U_{BC}^\dag)]\\
= &\ E_{A|B}[\Lambda_B(\rho_{AB})].
\end{aligned}
\end{equation}\hfill$\Box$

\textbf{Theorem 2.} The entanglement measure based on distance $E(\rho_{AB})$, which satisfies Eqs. (3), (5) and (6), must be maximal on pure states under operations on one side.

\emph{Proof}. Any state $\rho_{AB}$ on $\mathbb{C}^{d}\otimes\mathbb{C}^{d}$ can be seen as the result of the application of a channel $\Lambda_B$ ($\Lambda_A$) on any purification $|\psi\rangle_{AB}\langle\psi|\in\mathbb{C}^{d}\otimes\mathbb{C}^{d}$ of $\rho_{B}$($\rho_{A}$) [20, 21]. Then, $E_{A|B}(|\psi\rangle_{AB}\langle\psi|)\geq E_{A|B}[\Lambda_B(|\psi\rangle_{AB}\langle\psi|)]=E_{A|B}(\rho_{AB})$. Here the inequality is due to Theorem 1. \hfill$\Box$

Next we extend these properties to any bipartite entanglement measures, the following results are obtained.

\textbf{ Corollary 1.} A bipartite measure of entanglement $E$, which satisfies Eqs. (3), (5) and (6), must be maximal on pure states under operations on one side.

The same conclusions can be found in [21], where they arrive at this result  by utilizing the monogamy relation
\begin{align*}
E_{A|BC}(\rho_{A|BC})\geq E_{A|B}(\rho_{AB})+ E_{A|C}(\rho_{AC}).
\end{align*}
However, we get these results  by using the basic properties of the measures of entanglement based on distance.

In the following we consider two special  measures of entanglement based on distance. The first is Bures measure of entanglement, which can be written as [17, 18]
\begin{equation}\label{8}
 \begin{aligned}
 E_{\texttt{B}}(\rho_{AB})= \min_{\sigma\in S}(2-2\sqrt{F(\rho_{AB},\sigma_{AB})}).
\end{aligned}
\end{equation}
Here $F(\rho,\sigma)= (\text{tr}[\sqrt{\sqrt{\rho_{AB}}\sigma_{AB}\sqrt{\rho_{AB}}}])^2$ is the fidelity.

The second is the geometric measure of entanglement [19, 22], which is defined as
\begin{equation}\label{9}
 \begin{aligned}
 E_{\texttt{G}}(\rho_{AB})= \min_{\sigma\in S}(1-F(\rho_{AB},\sigma_{AB})).
\end{aligned}
\end{equation}

It has been proved that  for a two-qubit state, the Bures measure of entanglement as a function of the concurrence $C(\rho_{AB})$ has an analytical formula [22]
\begin{equation}\label{10}
 \begin{aligned}
 E_{\texttt{B}}(\rho_{AB})= B(C(\rho_{AB}));
\end{aligned}
\end{equation}
the geometric measure of entanglement for a two-qubit state as a function of the concurrence has  the analytical expression [19]
\begin{equation}\label{11}
 \begin{aligned}
 E_{\texttt{G}}(\rho_{AB})= G(C(\rho_{AB})).
\end{aligned}
\end{equation}
Here
\begin{equation}\label{12}
 \begin{aligned}
B(x)=2-2\sqrt{\frac{1+\sqrt{1-x^{2}}}{2}}
\end{aligned}
\end{equation}
and
\begin{equation}\label{13}
 \begin{aligned}
G(x)=\frac{1-\sqrt{1-x^{2}}}{2},
\end{aligned}
\end{equation}
 both $B(x)$ and $G(x)$ are monotonically increasing functions in $0\leq x\leq1$.

Let us recall the definition of concurrence. For a bipartite pure state $|\phi\rangle_{AB}$, the concurrence is given by [23]
\begin{equation}\label{14}
C(|\phi\rangle_{AB})=\sqrt{2(1-\text{tr}\rho_{A}^{2})},
\end{equation}
where $\rho_{A}=\text{tr}_{B}(|\phi\rangle_{AB}\langle\phi|)$. For a mixed state $\rho_{AB}$, the concurrence is defined via the convex-roof extension
\begin{equation}\label{15}
C(\rho_{AB})=\min\sum_{j}p_{j}C(|\phi_{j}\rangle_{AB}),
\end{equation}
where the minimum is taken over all possible pure-state decompositions of $\rho_{AB}=\sum_{j}p_{j}|\phi_{j}\rangle_{AB}\langle\phi_{j}|$.

For an arbitrary $N$-qubit state $\rho_{AB_{1}\cdots B_{N-1}}\in \mathcal{H}_{A}\otimes \mathcal{H}_{B_1}\otimes\cdots\otimes\ \mathcal{H}_{B_{N-1}}$, we use $\rho_{A|B_{1}\cdots B_{N-1}}$ to denote the state $\rho_{AB_{1}\cdots B_{N-1}}$ viewed as a bipartite state with partitions $A$ and $B_{1}B_{2}\cdots B_{N-1}$. Here $\mathcal{H}_{A}, \mathcal{H}_{B_1},\cdots, \mathcal{H}_{B_{N-1}}$ are two-dimensional Hilbert spaces of the systems $A, B_{1}, \cdots, B_{N-1}$, respectively. The concurrence $C(\rho_{A|B_{1}\cdots B_{N-1}})$ satisfies [6]
\begin{equation}\label{16}
C^{2}(\rho_{A|B_{1}\cdots B_{N-1}})-C^{2}(\rho_{AB_{1}})-\cdots-C^{2}(\rho_{AB_{N-1}})\geq0,
\end{equation}
where $\rho_{AB_{i}}=\text{tr}_{B_{1}\cdots B_{i-1}B_{i+1}\cdots B_{N-1}}(\rho_{A|B_{1}\cdots B_{N-1}})$.

In order to investigate the monogamy inequality for Bures measure of entanglement and the geometric measure of entanglement, we need the following lemma. Here we only present the result. The detailed proof of the result is given in Appendix.

\textbf{Lemma} For $\eta\geq 1$, we have
\begin{equation}\label{17}
 \begin{aligned}
B{^{\eta}}(\sqrt{x^{2}+y^{2}})\geq B{^{\eta}}(x)+B{^{\eta}}(y)
\end{aligned}
\end{equation}
and
\begin{equation}\label{18}
 \begin{aligned}
G{^{\eta}}(\sqrt{x^{2}+y^{2}})\geq G{^{\eta}}(x)+G{^{\eta}}(y)
\end{aligned}
\end{equation}
on the domain $D=\{(x,y)|0\leq x,y,x^{2}+y^{2}\leq1\}$.

We now analyze the monogamy relation in an $N$-qubit pure quantum state. According to the Schmidt decomposition, the subsystem $B_{1}\cdots B_{N-1}$ can be regarded as a logic qubit. Thus, the Bures measure of entanglement and the geometric measure of entanglement can be evaluated using Eq.(10) and Eq.(11) respectively. Then we can derive the following monogamy relation.
 \begin{equation}\label{19}
 \begin{aligned}
  E_{\xi}^{\eta}(|\Psi\rangle_{A|B_{1}\cdots B_{N-1}}) & =
  [ E_{\xi}(C(|\Psi\rangle_{A|B_{1}\cdots B_{N-1}}))]^{\eta}\\
  & \geq
  \left [ E_{\xi}\left (\sqrt{C^{2}(\rho_{AB_{1}})+\cdots+C^{2}(\rho_{AB_{N-1}})} \right )\right ]^{\eta}\\
  & \geq
  [ E_{\xi}(C(\rho_{AB_{1}}))]^{\eta}+\left [ E_{\xi}\left(\sqrt{C^{2}(\rho_{AB_{2}})+\cdots+C^{2}(\rho_{AB_{N-1}})}\right)\right]^{\eta}\\
  & \geq
  [ E_{\xi}(C(\rho_{AB_{1}}))]^{\eta}+[ E_{\xi}(C(\rho_{AB_{2}}))]^{\eta}
  +\cdots+[ E_{\xi}(C(\rho_{AB_{N-1}}))]^{\eta}\\
  & =
   E_{\xi}^{\eta}(\rho_{AB_{1}})+ E_{\xi}^{\eta}(\rho_{AB_{2}})+\cdots+E_{\xi}^{\eta}(\rho_{AB_{N-1}}),
\end{aligned}
\end{equation}
where $E_{\xi}\in\{E_{\texttt{B}}, E_{\texttt{G}}\}$. We have utilized the monogamy inequality (16) and the monotonically increasing property of the function $B(x)$ and $G(x)$ to obtain the first inequality, the second inequality is due to inequality (17) and (18) by letting $x=C(\rho_{AB_{1}})$ and $y=\sqrt{C^{2}(\rho_{AB_{2}})+\cdots+C^{2}(\rho_{AB_{N-1}})}$. The third inequality is obtained from the iterative use of inequality (17) and (18). Since for any two-qubit state $\rho_{AB}$, $E_{\xi}(\rho_{AB})= E_{\xi}(C(\rho_{AB}))$, we obtain the last equality.

Especially, we choose $\eta=1$, then the inequality (19) becomes
\begin{equation}\label{20}
 \begin{aligned}
E_{\texttt{B}}(|\Psi\rangle_{A|B_{1}\cdots B_{N-1}}) \geq E_{\texttt{B}}(\rho_{AB_{1}})+ E_{\texttt{B}}(\rho_{AB_{2}})+\cdots+E_{\texttt{B}}(\rho_{AB_{N-1}}),
\end{aligned}
\end{equation}
and
\begin{equation}\label{21}
 \begin{aligned}
  E_{\texttt{G}}(|\Psi\rangle_{A|B_{1}\cdots B_{N-1}}) \geq E_{\texttt{G}}(\rho_{AB_{1}})+ E_{\texttt{G}}(\rho_{AB_{2}})+\cdots+E_{\texttt{G}}(\rho_{AB_{N-1}}).
\end{aligned}
\end{equation}
Hence we have completed the proof showing that Bures measure of entanglement and the geometric measure of entanglement with a power $\eta$ obey a general monogamy relation in an arbitrary $N$-qubit pure state for $\eta\geq1$.

In the following, we establish a class of tight monogamy relations by using the power of Bures measure of entanglement and the geometric measure of entanglement. Let us begin by recalling the Lemma 2 and Lemma 3 in [24]. For $a_{1}\geq a_{2}\geq \cdots \geq a_{n}\geq0$, $\mu\geq1$, then
\begin{equation}\label{22}
(a_{1}+a_{2}+\cdots+a_{n})^{\mu} \geq a_{1}^{\mu}+(2^{\mu}-1)a_{2}^{\mu}+\cdots+[n^{\mu}-(n-1)^{\mu}]a_{n}^{\mu}.
\end{equation}
For $t\geq1$ and $0\leq x\leq \frac{1}{k}$ with $k\geq1$, then
\begin{equation}\label{23}
(1+x)^{t}\geq1+\frac{k t}{k+1} x+[(k+1)^{t}-(1+\frac{t}{k+1})k^{t}]x^{t}.
\end{equation}

We now present a general framework for the monogamy relations for pure states of multiqubit systems.

\textbf{Theorem 3}. For an arbitrary $N$-qubit pure state $|\Psi\rangle_{A|B_{1}\cdots B_{N-1}}\in \mathcal{H}_{A}\otimes \mathcal{H}_{B_{1}}\otimes\cdots\otimes \mathcal{H}_{B_{N-1}}$, we can always have $E_{\xi}(\rho_{A|B_{i}})\geq E_{\xi}(\rho_{A|B_{i+1}})$ for $i=1,2,\cdots,N-2$, $N\geq 3$ by relabeling the subsystems (if necessary). For $\eta\geq1$, we have the monogamy relation as
\begin{equation}\label{24}
 \begin{aligned}
 E_{\xi}^{\eta}(|\Psi\rangle_{A|B_{1}\cdots B_{N-1}}) \geq
 E_{\xi}^{\eta}(\rho_{A|B_{1}})+(2^{\eta}-1)E_{\xi}^{\eta}(\rho_{A|B_{2}})+\cdots
 +[(N-1)^{\eta}-(N-2)^{\eta}]E_{\xi}^{\eta}(\rho_{A|B_{N-1}}),
\end{aligned}
\end{equation}
where $E_{\xi}\in\{E_{\texttt{B}}, E_{\texttt{G}}\}$.

\emph{Proof}. From the inequalities (20) and (21), if $\eta\geq1$, then we arrive at
\begin{equation}\label{25}
 \begin{aligned}
 E_{\xi}^{\eta}(|\Psi\rangle_{A|B_{1}\cdots B_{N-1}}) \geq \left[E_{\xi}(\rho_{AB_{1}})+ E_{\xi}(\rho_{AB_{2}})+\cdots+E_{\xi}(\rho_{AB_{N-1}})\right]^{\eta}.
\end{aligned}
\end{equation}

When $E_{\xi}(\rho_{A|B_{i}})\geq E_{\xi}(\rho_{A|B_{i+1}})$ for $i=1,2,\cdots,N-2$, $N\geq 3$, by the inequality (22), there is
\begin{equation}\label{26}
 \begin{aligned}
 E_{\xi}^{\eta}(|\Psi\rangle_{A|B_{1}\cdots B_{N-1}}) \geq
 E_{\xi}^{\eta}(\rho_{A|B_{1}})+(2^{\eta}-1)E_{\xi}^{\eta}(\rho_{A|B_{2}})+\cdots
 +[(N-1)^{\eta}-(N-2)^{\eta}]E_{\xi}^{\eta}(\rho_{A|B_{N-1}}).
\end{aligned}
\end{equation}

Next, we show that the monogamy relations in Theorem 3 can even be further improved to be tighter under certain conditions on the Bures measure of entanglement and the geometric measure of entanglement.

\textbf{Theorem 4.}  For an arbitrary $N$-qubit pure state $|\Psi\rangle_{AB_{1}\cdots B_{N-1}}\in \mathcal{H}_{A}\otimes \mathcal{H}_{B_{1}}\otimes\cdots\otimes \mathcal{H}_{B_{N-1}}$, if $E_{\xi}(\rho_{A|B_{i}})\geq k\sum\limits_{l=i+1}^{N-1} E_{\xi}(\rho_{A|B_{l}})$ for $i=1, 2, \cdots, m,$ and $k' E_{\xi}(\rho_{A|B_{j}})\leq\sum\limits_{l=j+1}^{N-1} E_{\xi}(\rho_{A|B_{l}})$ for $j=m+1, \cdots, N-2, \forall\ 1\leq m\leq N-3, N\geq4$, then
\begin{equation}\label{27}
 \begin{aligned}
E_{\xi}^{\eta}(|\Psi\rangle_{A|B_{1}\cdots B_{N-1}}) \geq &
E_{\xi}^{\eta}(\rho_{A|B_{1}})+[(k+1)^{\eta}-k^{\eta}]E_{\xi}^{\eta}(\rho_{A|B_{2}})+\cdots
 +[(k+1)^{\eta}-k^{\eta}]^{m-1}E_{\xi}^{\eta}(\rho_{A|B_{m}})\\
& +[(k+1)^{\eta}-k^{\eta}]^{m}[(k'+1)^{\eta}-k'^{\eta}][E_{\xi}^{\eta}(\rho_{A|B_{m+1}})+\cdots+E_{\xi}^{\eta}(\rho_{A|B_{N-3}})]\\
& +\left[(k+1)^{\eta}-k^{\eta}\right]^{m}\left\{\left[(k'+1)^{\eta}-(1+\frac{\eta}{k'+1})k'^{\eta}\right]E_{\xi}^{\eta}(\rho_{A|B_{N-2}})\right.\\
& +\left.\frac{k' \eta}{k'+1}E_{\xi}(\rho_{A|B_{N-2}})E_{\xi}^{\eta-1}(\rho_{A|B_{N-1}})+E_{\xi}^{\eta}(\rho_{A|B_{N-1}})\right\}
\end{aligned}
\end{equation}
for $\eta\geq 1$, $k\geq1$, $k'\geq1$, where $E_{\xi}\in\{E_{\texttt{B}}, E_{\texttt{G}}\}$.

\emph{Proof}. From the inequalities (20) and (21), we can derive
\begin{equation}\label{28}
 \begin{aligned}
 E_{\xi}^{\eta}(|\Psi\rangle_{A|B_{1}\cdots B_{N-1}}) & \geq
 E_{\xi}^{\eta}(\rho_{A|B_1})+\frac{k \eta}{k+1}E_{\xi}^{\eta-1}(\rho_{A|B_1})\left(\sum\limits_{l=2}^{N-1}E_{\xi}(\rho_{A|B_{l}})\right)\\
   & \quad+\left[(k+1)^{\eta}-(1+\frac{\eta}{k+1})k^{\eta}\right]\left(\sum\limits_{l=2}^{N-1}E_{\xi}(\rho_{A|B_{l}})\right)^{\eta}\\
  & \geq E_{\xi}^{\eta}(\rho_{A|B_{1}})+ [(k+1)^{\eta}-k^{\eta}]E_{\xi}^{\eta}(\rho_{A|B_{2}})+\cdots+[(k+1)^{\eta}-k^{\eta}]^{m-2}E_{\xi}^{\eta}(\rho_{A|B_{m-1}})\\
  & \quad +[(k+1)^{\eta}-k^{\eta}]^{m-1}\left[E_{\xi}^{\eta}(\rho_{A|B_{m}})+\frac{k \eta}{k+1}E_{\xi}^{\eta-1}(\rho_{A|B_{m}})\left(\sum\limits_{l=m+1}^{N-1}E_{\xi}(\rho_{A|B_{l}})\right)\right.\\
  & \quad+\left.[(k+1)^{\eta}-(1+\frac{\eta}{k+1})k^{\eta}]\left(\sum\limits_{l=m+1}^{N-1}E_{\xi}(\rho_{A|B_{l}})\right)^{\eta}\right],
\end{aligned}
\end{equation}
where the first inequality follows from the inequality (23). The second inequality is obtained from the iterative use of inequality (23). It is noted that we also exploit the fact that $1+\frac{k t}{k+1} x+[(k+1)^{t}-(1+\frac{t}{k+1})k^{t}]x^{t}\geq 1+[(k+1)^{t}-k^{t}]x^{t}$ for $0\leq x\leq \frac{1}{k}, k\geq1$, $t \geq 1$, and $E_{\xi}(\rho_{A|B_{i}})\geq k \sum\limits_{l=i+1}^{N-1}E_{\xi}(\rho_{A|B_{l}})$, $i=1, 2, \cdots, m$.

When $k'E_{\xi}(\rho_{A|B_{j}})\leq \sum\limits_{l=j+1}^{N-1}E_{\xi}(\rho_{A|B_{l}})$ for $j=m+1, \cdots, N-2$, we can apply the preceding procedure to get
\begin{equation}\label{29}
 \begin{aligned}
\left(\sum\limits_{l=m+1}^{N-1}E_{\xi}(\rho_{A|B_{l}})\right)^{\eta} & \geq
\left[(k'+1)^{\eta}-(1+\frac{\eta}{k'+1})k'^{\eta}\right]E_{\xi}^{\eta}(\rho_{A|B_{m+1}})\\
& \quad+\frac{k' \eta}{k'+1}E_{\xi}(\rho_{A|B_{m+1}})\left(\sum\limits_{l=m+2}^{N-1}E_{\xi}(\rho_{A|B_{l}})\right)^{\eta-1}
+\left(\sum\limits_{l=m+2}^{N-1}E_{\xi}(\rho_{A|B_{l}})\right)^{\eta}\\
& \geq
[(k'+1)^{\eta}-k'^{\eta}][E_{\xi}^{\eta}(\rho_{A|B_{m+1}})+\cdots+E_{\xi}^{\eta}(\rho_{A|B_{N-3}})]\\
& \quad+\left[(k'+1)^{\eta}-(1+\frac{\eta}{k'+1})k'^{\eta}\right]E_{\xi}^{\eta}(\rho_{A|B_{N-2}})+\frac{k' \eta}{k'+1}E_{\xi}(\rho_{A|B_{N-2}})E_{\xi}^{\eta-1}(\rho_{A|B_{N-1}})\\
& \quad+E_{\xi}^{\eta}(\rho_{A|B_{N-1}}).
\end{aligned}
\end{equation}
Here we make use of the fact that $1+\frac{k' t}{k'+1} x+[(k'+1)^{t}-(1+\frac{t}{k'+1})k'^{t}]x^{t}\geq 1+[(k'+1)^{t}-k'^{t}]x^{t}$ for $0\leq x\leq \frac{1}{k'}, k'\geq1$, $t \geq 1$. Inequality (28) together with inequality (29) leads to inequality (27).  So we get Theorem 4.

We note that if $E_{\xi}(\rho_{A|B_{i}})\geq \sum\limits_{l=i+1}^{N-1} E_{\xi}(\rho_{A|B_{l}})$ for $i=1, 2, \cdots, m$, and $E_{\xi}(\rho_{A|B_{j}})\leq \sum\limits_{l=j+1}^{N-1} E_{\xi}(\rho_{A|B_{l}})$ for $j=m+1, \cdots, N-2, \forall\ 1\leq m\leq N-3, N\geq4$, then $k=1$, $k'=1$, we can get
\begin{equation}\label{30}
 \begin{aligned}
 E_{\xi}^{\eta}(|\Psi\rangle_{A|B_{1}\cdots B_{N-1}}) \geq &
 E_{\xi}^{\eta}(\rho_{A|B_{1}})+(2^{\eta}-1)E_{\xi}^{\eta}(\rho_{A|B_{2}})+\cdots
 +(2^{\eta}-1)^{m-1}E_{\xi}^{\eta}(\rho_{A|B_{m}})\\
& +(2^{\eta}-1)^{m+1}[E_{\xi}^{\eta}(\rho_{A|B_{m+1}})+\cdots+E_{\xi}^{\eta}(\rho_{A|B_{N-3}})]\\
& +(2^{\eta}-1)^{m}\left\{(2^{\eta}-\frac{\eta}{2}-1)E_{\xi}^{\eta}(\rho_{A|B_{N-2}})\right.\\
& +\left.\frac{\eta}{2}E_{\xi}(\rho_{A|B_{N-2}})E_{\xi}^{\eta-1}(\rho_{A|B_{N-1}})+E_{\xi}^{\eta}(\rho_{A|B_{N-1}})\right\}.
\end{aligned}
\end{equation}

In conclusions, we have shown that any entanglement measures based on distance must be maximal on pure states by basic properties of the entanglement measures based on distance. In particular, we have not only proven that Bures measure of entanglement and geometric measure of entanglement satisfy the monogamy inequality for multiqubit pure states, but also provided a class of tight monogamy relations for pure states of multiqubit systems for $\eta>1$ by using the power of Bures measure of entanglement and geometric measure of entanglement. The results provide a characterization of multipartite entanglement sharing and distribution for pure states of multi-qubit systems. We hope that the results presented in this paper are useful for the monogamy properties of the multipartite quantum entanglement and fully understood of the multipartite quantum entanglement.

\section*{APPENDIX: PROOF OF THE LEMMA }
\setcounter{equation}{0}
\renewcommand\theequation{A.\arabic{equation}}
 It is evident that the following inequalities
 \begin{equation}\label{A.1}
 \begin{aligned}
 \sqrt{1-x^{2}}+\sqrt{1-y^{2}}\geq 1+\sqrt{1-x^{2}-y^{2}}
 \end{aligned}
 \end{equation}
 and
 \begin{equation}\label{A.2}
 \begin{aligned}
 \sqrt{(1-x^{2})(1-y^{2})}\geq \sqrt{1-x^{2}-y^{2}}
 \end{aligned}
 \end{equation}
hold on the domain $D=\{(x,y)|0\leq x,y,x^{2}+y^{2}\leq1\}$. By using the above inequalities, we can obtain the following conclusion.

\textbf{Lemma} For $\eta\geq1$, we have
\begin{equation}\label{A.3}
 \begin{aligned}
B{^{\eta}}(\sqrt{x^{2}+y^{2}})\geq B{^{\eta}}(x)+B{^{\eta}}(y)
\end{aligned}
\end{equation}
and
\begin{equation}\label{A.4}
 \begin{aligned}
G{^{\eta}}(\sqrt{x^{2}+y^{2}})\geq G{^{\eta}}(x)+G{^{\eta}}(y)
\end{aligned}
\end{equation}
on the domain $D=\{(x,y)|0\leq x,y,x^{2}+y^{2}\leq1\}$.

\emph{Proof}. Using the inequality (A.1), it is easy to verify that
\begin{equation}\label{A.5}
\begin{aligned}
\frac{1-\sqrt{1-x^{2}-y^2}}{2}\geq\frac{1-\sqrt{1-x^{2}}}{2}+\frac{1-\sqrt{1-y^{2}}}{2}.
\end{aligned}
\end{equation}
Then we  get
\begin{equation}\label{A.6}
 \begin{aligned}
G(\sqrt{x^{2}+y^{2}})\geq G(x)+G(y).
\end{aligned}
\end{equation}

When $\eta\geq 1$, by the inequality (A.6), one derives
\begin{equation}\label{A.7}
\begin{aligned}
\left[\frac{1-\sqrt{1-x^{2}-y^2}}{2}\right]^\eta & \geq \left[\frac{1-\sqrt{1-x^{2}}}{2}+\frac{1-\sqrt{1-y^2}}{2}\right]^\eta\\
& \geq \left[\frac{1-\sqrt{1-x^{2}}}{2}\right]^\eta+\left[\frac{1-\sqrt{1-y^2}}{2}\right]^\eta,
\end{aligned}
\end{equation}
where in the second inequality we have used the property $(1+x)^\eta\geq 1+x^\eta$ for $0\leq x\leq1$ and $\eta\geq 1$. It implies that the inequality (A.4) holds, and completes the proof of the inequality (A.4).

Using the inequalities (A.1) and (A.2), we have
\begin{equation}\label{A.8}
\begin{aligned}
(1+\sqrt{1-x^{2}})(1+\sqrt{1-y^{2}})\geq 2+2\sqrt{1-x^{2}-y^2}.
\end{aligned}
\end{equation}
Inequalities (A.5) and (A.8) together yield
\begin{equation}\label{A.9}
\begin{aligned}
1+\frac{\sqrt{1-x^{2}}}{2}+\frac{\sqrt{1-y^2}}{2}+\sqrt{(1+\sqrt{1-x^{2}})(1+\sqrt{1-y^{2}})}\geq 1+\frac{1+\sqrt{1-x^{2}-y^{2}}}{2}+\sqrt{2+2\sqrt{1-x^{2}-y^2}}.
\end{aligned}
\end{equation}
We can write (A.9) as
\begin{equation}\label{A.10}
\begin{aligned}
 \sqrt{\frac{1+\sqrt{1-x^{2}}}{2}}+ \sqrt{\frac{1+\sqrt{1-y^{2}}}{2}}\geq 1+ \sqrt{\frac{1+\sqrt{1-x^{2}-y^2}}{2}}.
 \end{aligned}
\end{equation}
It follows that
\begin{equation}\label{A.11}
\begin{aligned}
2-2\sqrt{\frac{1+\sqrt{1-x^{2}-y^2}}{2}}\geq 2-2\sqrt{\frac{1+\sqrt{1-x^{2}}}{2}}+2-2\sqrt{\frac{1+\sqrt{1-y^2}}{2}},
 \end{aligned}
\end{equation}
that is
\begin{equation}\label{A.12}
 \begin{aligned}
B(\sqrt{x^{2}+y^{2}})\geq B(x)+B(y).
\end{aligned}
\end{equation}
When $\eta\geq 1$, the proof of inequality (A.3) is similar to the proof of the inequality (A.4). It is now obvious that the lemma holds.

\vspace{0.6cm}
\acknowledgments
This work was supported by the National Natural Science Foundation of China under Grant No: 11475054, the Hebei Natural Science Foundation of China under Grant No:  A2018205125.

\end{document}